\documentclass[usenatbib]{mn2e}
\usepackage{graphicx}
\usepackage{astrobib_mnras2e}

\newcommand{\beq}{\begin{equation}}
\newcommand{\eeq}{\end{equation}}
\newcommand{\beqa}{\begin{eqnarray}}
\newcommand{\eeqa}{\end{eqnarray}}

\newcommand\kev{{\rm\ keV}}

\begin{document}

\title{Luminous Red Galaxy Population in Clusters at $0.2\le z\le 0.6$}
\author[Ho et al.]
{Shirley Ho${}^1$\thanks{shirley@astro.princeton.edu},
Yen-Ting Lin${}^{1,2}$, David Spergel$^{1}$ and Christopher M. Hirata$^{3}$\\
$^{1}$ Department of Astrophysical Sciences, Peyton Hall,
Princeton University, Princeton, NJ 08544, USA.\\
$^{2}$ Departamento de Astronom\'{i}a y Astrof\'{i}sica, Pontificia Universidad Cat\'{o}lica de Chile, Chile.\\
$^{3}$School of Natural Sciences, Institute for Advanced Study, Princeton, NJ 08540, USA.}

\date{\today}

\maketitle
\begin{abstract}
We investigate statistical properties of LRGs in a sample of X-ray
selected galaxy clusters at intermediate redshift ($0.2\le z\le0.6$).
The LRGs are selected based on carefully designed color criteria, and
the cluster membership is assessed via photometric redshifts. 
As clusters and LRGs are both viewed as promising tracer of the underlying
dark matter distribution, understanding the distribution of LRGs within
clusters is an important issue.
Our main findings include:

1. The halo occupation distribution of LRGs inside our
cluster sample is  $N(M) = k\times (M/10^{14})^{a}$ where 
$a=0.620\pm 0.105 $ and $k=1.425\pm0.285 $ assuming a Poisson distribution
for $N(M)$.

2. The halo occupation distribution of LRGs ($N(M)$)  and 
   the satellite distribution of LRGs ($N-1(M)$) are both consistent with being Poisson.
   To be more quantitative, we find $Var(N)/\langle N\rangle= 1.428\pm 0.351$ and 
   $Var(N-1)/\langle N-1\rangle = 1.823 \pm 0.496$. 

3. The radial profile of LRGs within clusters when fitted with
a NFW profile gives a concentration of $17.5^{+7.1}_{-4.3}$ ($6.0^{+3.2}_{-1.9}$) including (excluding) BLRGs (Brightest LRGs).

We also discuss the implications of these observations on the evolution of massive galaxies in clusters.
\end{abstract}

\section{Introduction}
\label{sec:intro}

The recent advent of large-scale galaxy surveys have revolutionized the
field of observational cosmology. 
Deep spectroscopic surveys allows us to witness
the young Universe when the building blocks of present-day galaxies are
forming, and some of the well-known properties of the local galaxy populations
are about to realize. The enormous amount of data gathered by wide area 
surveys produce galaxy samples with exquisite statistical precision, which makes
it possible to single out the most fundamental properties that govern the physics 
of galaxy formation from the medley of observables.

Equally impressive has been the progress in the theoretical understanding
of the structure formation in the Universe. Techniques such as direct numerical
simulations and semi-analytic models can now reproduce the observed 
properties of galaxies, such as the luminosity function and 2-point correlation function, color, mass-to-light ratios 
over large ranges of environments and cosmic epochs \citep{kauffmann99a,kauffmann99b,springel01b,cole00,swhite78}.

Yet another approach, the so-called halo model, which is phenomenological in 
nature, has enjoyed popularity over the recent years. An essential ingredient
of this method is the halo occupation distribution (HOD), which refers to the
way galaxies (or substructures of dark matter halos) ``populate'' dark matter halos.
In general, an HOD description includes the mean number of galaxies per halo 
$N$ as a function of halo mass, the probability distribution that a halo of mass
$M$ contains $N$ galaxies $P(N|M)$, and the relative distribution (both spatial 
and velocity) of galaxies and dark matter within halos \citep{berlind02}.

The halo model formalism allows fast exploration of a wide range of HODs; an
HOD that reproduces the observed clustering properties and luminosity function
of galaxies can be further studied to reveal the physical processes that lead to
galaxy formation and understanding of cosmological parameters. 
Examples of using halo model formalism to reproduce observables in order 
to reveal parameters in cosmology, galaxy evolution and formation includes 
\citep[e.g.][]{abazajian05,white07,yoo06,zheng07,kulkarni07}.

Despite the success in both observational and theoretical sides, there remains 
some unsolved problems regarding the formation of the massive, (usually) early 
type, galaxies. These galaxies appear '' red and dead'', with the majority of the 
stars forming at high redshift ($z > \sim 2$) and evolving passively since. 
Within the cold dark matter (CDM) paradigm, in which massive galaxies are built
by smaller galaxies via mergers in the late times, mergers between gas poor 
systems (``dry'' mergers) seem to be a promising route to form giant galaxies. 
Observationally, however, the overall importance of dry mergers is still under 
heated debate.


Luminous Red Galaxies (LRGs) are massive galaxies composed mainly
of old stars, with little or no on-going star formation.
They demonstrate very consistent spectral energy distribution (SED). Their SEDs mainly
consist of old star spectrum, most notably for the $4000$\AA\ break. This allows one to 
photometrically determine their redshifts fairly accurately \citep[see ][]{padmanabhan05}. 
With the accurate photometric redshifts of LRGs, one can probe a larger volume of the universe, thus
giving better constraints on the formation of massive galaxies.
By studying the HOD of the LRGs, we aim to provide a simple quantitative
description of these galaxies in massive dark matter halos, which will enable
direct comparison with predictions of galaxy formation models.

Here we aim to provide observational constraints on the HOD of the LRGs
based on a sample of 47 intermediate-redshift clusters from the ROSAT 400d survey
\citep{burenin06}, with photometric data from the
Sloan Digital Sky Survey \citep[SDSS;][]{stoughton02}. Using X-ray properties of
these clusters to define the cluster center and estimate the cluster binding
mass, we determine the mean halo occupation number $N$ as a function of mass
from $\sim 1 \times 10^{14} M_\odot$ to $\sim 8 \times 10^{14} M_\odot$ and also
investigate the LRG distribution and luminosity distribution within the clusters.

In \S\ref{sec:Data}, we briefly describe the
X-ray cluster catalog that we utilize and
the construction of SDSS LRG sample.
In \S\ref{sec:Analysis}, we present our method and findings on
the LRG distributions within the clusters and the 
mean halo occupation number.
We discuss what is a  good mass tracers and evolution of massive galaxies
in 
\S\ref{sec:discussion}.  Possible systematics that may affect our results are
discussed in \S\ref{sec:system}. 


Throughout the paper we assume the cosmological parameters to be 
the WMAP values \citep{spergel06}: $\Omega_m h^2 = 0.1277$,
$h = 0.732$ and the Hubble
parameter $H_0=73\,h_{73}$~km~s$^{-1}$~Mpc$^{-1}$.

%

\section{Data}
\label{sec:Data}
\subsection{Cluster Sample}

Our cluster sample is drawn from the 400 square degree ROSAT PSPC Galaxy
Cluster Survey \citep{burenin06} (hereafter the 400d survey), which is an extension
of the 160 square degree survey \citep{vikhlinin98}. The survey detects extended
X-ray sources in archival ROSAT PSPC images down to a flux limit of $1.4\times 10^{-13}\,
{\rm erg\,s}^{-1} {\rm cm}^{-2}$, with extensive optical spectroscopic follow up.
Out of the 266 clusters detected in the survey, 47 lie within the redshift range
$0.2\le z\le 0.6$ and are covered by SDSS DR5. The redshift range is
chosen to be consistent with the photometric cuts designed to select a homogeneous
LRG sample across a wide range in cosmic epochs.

The cluster catalog from the 400d survey provides estimates of cluster center, redshift,
and X-ray luminosity $L_X$, which is used to estimate the cluster mass. Some of the
basic information of the clusters in our sample is given in Table~1.

\subsection{LRG Data from Sloan Digital Sky Survey}
\label{sec:LRGdata}

%
%
The Sloan Digital Sky Survey has taken $ugriz$ CCD images of $10^4$ deg$^2$ of the high-latitude
sky.
A dedicated 2.5m telescope at Apache Point Observatory images the sky in 5 bands between
3000\AA\  and 10000\AA\ \citep{fukugita96}
using a drift-scanning, mosaic CCD camera \citep{gunn98,gunn06}, detecting
objects to a flux limit of $r\sim 22.5$ mag.
The survey selects $10^6$ targets for spectroscopy, most of them galaxies with
$r < 17.77$ mag \citep{gunn98, york00, stoughton02}. This spectroscopic
follow-up uses two digital spectrographs on the same telescope as the imaging camera.
Details of the galaxy survey are described in the galaxy target selection papers \citep{eisenstein01,strauss02}; other aspects of the survey are mainly described
in the Early Data Release paper \citep{stoughton02}.
All the data processing, including astrometry \citep{pier03}, source
identification and photometry \citep{lupton01,hogg01,ivezic04}, 
calibration \citep{fukugita96,smith02}, spectroscopic target selection
\citep{eisenstein01,strauss02,richards02}, and spectroscopic fiber placement 
\citep{blanton03} are done automatically via SDSS software \citep{tucker06}.
The SDSS is well-underway, and has had six major releases  
\citep{Adel07}.

\begin{figure}
\begin{center}
\includegraphics[width=3.0in]{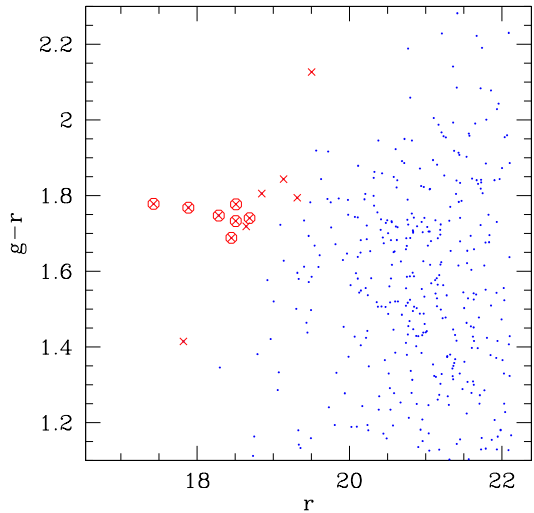}
\end{center}
\caption{Color-magnitude diagram of objects toward the field of Cluster 142. Those satisfying cuts I and II (see Eqn.~\ref{eq:colorcuts} and \ref{eq:magcuts}) are shown as crosses and squares respectively. The red circles are selected LRGs and the blue dots are objects detected in SDSS photometrically. As shown, the selected LRGs lie very systematically along the red sequence of the cluster. }
\label{fig:142gr}
\end{figure}

We utilize the photometric LRGs from SDSS
constructed as described in \citet[][hereafter P05]{padmanabhan05}.
The LRGs have been very useful as a cosmological probe since they are typically the most luminous
galaxies in the universe, thus they probe a larger volume than most other tracers. On top of this,
they also have very regular spectral energy distributions and a prominent 4000\AA\ break, making photometric redshift estimation much easier than the other galaxies.
We plot the color magnitude diagram for one of the cluster and show that the LRGs in the cluster are the bright red galaxies 
that follow nicely along the red sequence (see Fig~\ref{fig:142gr}). 

Our selection criteria are based on the spectroscopic
selection of LRGs described in \citet{eisenstein01},
extended to lower apparent luminosities (P05). We
select LRGs by choosing galaxies that both have colors consistent with
an old stellar population, as well as absolute luminosities greater
than a chosen threshold. The first criterion is simple to implement since
the uniform SEDs of LRGs imply that they lie on an extremely tight locus
in the space of galaxy colors; we simply select all galaxies that lie
close to that locus. More specifically, we can define three (not
independent) colors that describe this locus,
\begin{eqnarray}
c_{\perp} &\equiv & (r-i) - 0.25(g-r) - 0.18 \,\,\, , \nonumber \\
d_{\perp} &\equiv & (r-i) - 0.125(g-r) \,\,\,, \nonumber \\
c_{||} &\equiv & 0.7 (g-r) + 1.2(r-i-0.18) \,\,\,,
\label{eq:perpdef}
\end{eqnarray}
where $g$, $r$, and $i$ are the SDSS model magnitudes
in these bands respectively.
We now make the following color selections,
\begin{eqnarray}
{\rm Cut\,\,I :} & \mid c_{\perp} \mid < 0.2; \nonumber \\
{\rm Cut\,\,II :} & d_{\perp} > 0.55, \,\,\, g-r > 1.4,
\label{eq:colorcuts}
\end{eqnarray}
%
as well as the magnitude cuts
\begin{eqnarray}
{\rm Cut\,\,I :} & r_{Petro} < 13.6 + c_{||}/0.3; \nonumber \\
                 & r_{Petro} <19.7; \nonumber \\
{\rm Cut\,\,II :} & i < 18.3 + 2 d_{\perp}, \nonumber \\ 
                  & i< 20.
\label{eq:magcuts}
\end{eqnarray}

Making two cuts (Cut I and Cut II) is convenient since the LRG
color locus changes direction sharply as the 4000\AA\ break
redshifts from the $g$ to the $r$ band; this division divides the
sample into low redshift (Cut I, $z < 0.4$) and high
redshift (Cut II, $z > 0.4$) samples.
More details of these color selection criteria are thoroughly described in
P05.

We do however apply slightly different cuts than those adopted in
P05: we limit our samples to sky regions where
\begin{equation}
E(B-V) \le 0.08
\end{equation}
and data taken under seeing condition of
\begin{equation}
FWHM < 2.0''.
\end{equation}
These cuts in extinction and seeing are applied simply by excluding areas at which
the galaxy overdensity drops significantly. 
Furthermore, there are a few regions in SDSS that have 60\% more red objects
and less blue objects; we decide to throw away these regions.

We slice our LRG sample into two redshift bins: $0.2\le z_{photo}\le 0.4$ and  
$0.4\le z_{photo} \le0.6$.
We also regularized our redshift distribution as described in P05.
For our sample, we have  855534 galaxies, covering 2,025,731 resolution 10 
HEALpix pixels, each with area of $11.8$ $\rm {arcmin}^2$, giving 0.422334 gal/pix.

We then estimate the photometric redshift of these LRGs with the algorithm developed
by P05. The typical uncertainty of the photo-$z$'s is 
$\delta_z = \sigma_z/(1+z)\approx 0.03$ (see P05).

\section{Analysis}
\label{sec:Analysis}

\subsection{Method}
\label{sec:method}
We estimate the cluster virial mass $M_{200} \equiv (4\pi/3)r_{200}^3\times 200\rho_c$ 
from the X-ray luminosity using the mass--luminosity relation given by \citet{reiprich02}
\begin{equation} 
\label{eq:ra_xlm} 
\log \left[{1.4^2L_X(0.1-2.4\kev) \over h_{73}^{-2} 10^{40}\, {\rm erg\ s}^{-1}}
\right] = A + \alpha \log \left( {1.4 M_{200} \over h_{73}^{-1} M_\odot} \right), 
\end{equation}  
where $A=-20.055$ and $\alpha=1.652$. The radius $r_{200}$ is defined such that the
enclosed mean overdensity is 200 times the critical density $\rho_c$. The corresponding
angular extent is $\theta_{200}$. The mass--luminosity
scaling relation provides a mass estimate accurate to $< 50\%$
\citep{reiprich02} and a virial radius $r_{200}$ estimate accurate to 15\%.

As we now have the redshifts and positions of these clusters, we locate the LRGs as described in 
\S \ref{sec:LRGdata} 
in each of these clusters. 
We look for LRGs that are within a cylinder of radius $\theta_{200}$ and length of $\Delta_{z} = 0.06$ from the cluster center in both position and 
redshift space (i.e. $z_{LRG} = z_c \pm 0.03$). We choose $\delta_{z} = 0.03$ since that is the typical $1\sigma$ error 
on the LRG photometric redshift (P05) and $z_c$ is the cluster redshift. 
More discussion on the choice of cluster radius
and $\delta_{z}$ will be described in \S\ref{sec:system}.

Since we are relying on the photometric redshifts of the LRGs to find out whether 
a LRG sits in certain cluster or not, we take into account the effects of the following 
mechanisms that may lead to over-(or under-)estimate of the number of LRGs in each cluster:

i. LRG identification failure: There is an identification failure rate of $\sim 1\%$ \citep{padmanabhan06}. This is the rate of which a LRG (photometrically chosen) is actually a star or a quasar after 
we get the spectra of the object.


ii. Interlopers: There is a finite probability of finding LRGs inside the cluster purely by chance, we call these interlopers. 
We access the expected number of interlopers in each cluster
by looking at the average number of LRGs in sky (2D projected) in the 
solid angle of radius = $\theta_{200}$ of the cluster and the average probability
of finding a LRG in redshift range of $z_c\pm \delta_z $
where $\delta_z = 0.03$ (as defined above).
We can write down the expected number of interlopers ($\langle N_{int} \rangle $) as: 
\begin{equation}
\langle N_{int}\rangle = \bar{n} \pi \theta^2 \int_{z_c -\delta_z}^{z_c+ \delta_z} P(z_{p}) d z_{p}
\end{equation}
where $P(z_{p})$ is the normalized (photometric) redshift distribution of 
LRGs, $\bar{n}$ is the 2D average LRG density. 

iii. Missing galaxies due to errors in photometric redshift:
As an LRG can be scattered out of the cluster (due 
to photoz error), we need to account for this process by looking at the probability of 
LRG having been photometrically determined to be outside of the cluster, but 
in fact has spectroscopic redshift that falls within the range of the cluster:
\begin{equation}
P(|z_p-z_c|> \delta_z, | |z_s-z_c| < \delta_z) = \int_{z_{min}}^{z_{max}} (F(z_s) + B(z_s)) dz_s \,\,\, , \\
\end{equation}
\begin{eqnarray}
z_{min} = max(0.05,z_c-\delta_{z_c,in}) \,\,\, , \nonumber \\
z_{max} = min(0.7,z_c + \delta_{z_s,in}) \,\,\, , \nonumber \\
F(z) = \int_{-\infty}^{z_c-\delta_z-z} P(\delta,z)  d\delta  \,\,\, , \nonumber \\
B(z) = \int_{z_c+\delta_z-z}^{+\infty} P(\delta,z)  d\delta  \,\,\, ,
\end{eqnarray}
where $P(\delta,z)$ is the probability of finding $\delta$ ($= z_s - z_p$) at $z_s$, given
by \citet{padmanabhan05} and these are only characterized within the 
spectroscopic redshift range from $z=0.05$ to $z=0.7$. $\delta_{z_c,in}$ 
is the redshift range we allow a LRG to be a cluster member when 
we have its spectroscopic redshift, and this is set to be $0.01$. 

We then calculated the corrected LRG counts in each cluster via the following:
\begin{equation}
\langle N_{corr}\rangle = (\langle N_{obs}\rangle - \langle N_{int}\rangle) / f(z_p,z_c,z_s) \,\,\, , \\ 
\end{equation}
\begin{equation}
f(z_p,z_c,z_s) = (1-P(|z_p-z_c|> \delta_z | |z_s-z_c| < \delta_z) )\times (1+F)) \,\,\, , 
\end{equation}
and 
$F$ is the LRG identification failure rate. 

We list these corrected LRG counts in Table~1.

To convert the observed magnitudes of the LRGs into the rest-frame luminosity at $z=0$,
we follow the evolution of a simple stellar population formed in a burst at $z=5$,
with solar metallicity and Salpeter initial mass function, using the model of \citet{bruzual03}.
The LRGs are selected so that their present-day magnitude lies in the range $-23.5\le M_g\le -21$
(roughly corresponding to $1$--$7L_*$, where $L_*$ is the characteristic luminosity).

\begin{figure}
\begin{center}
\includegraphics[width=3.0in]{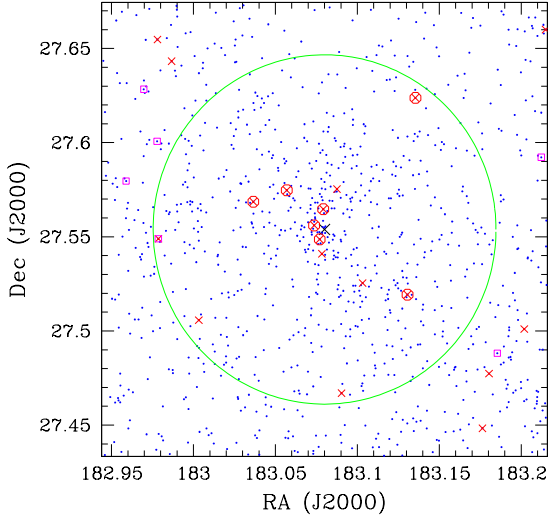}
\end{center}
\caption{Spatial distribution of LRGs in Cluster 142. The blue points represent all objects detected
in the SDSS photometric survey. Those satisfying cuts I and II are shown as crosses and squares,
respectively. The LRGs that have photo-$z$ consistent with the cluster redshift ($z=0.353$) are
represented as circles. The black cross denotes the centroid position of the ICM. The green circle
is the region encircled by the virial radius of the cluster.}
\label{fig:142}
\end{figure}

For each cluster, we visually inspect the spatial and color distributions of LRGs with respect to all 
objects detected by SDSS. An example is shown for cluster 142. Perhaps not surprisingly,
the spatial distribution of the LRGs seems concentrated towards cluster center (Fig.~\ref{fig:142}).

A general scenario that has been painted about LRGs and clusters is that
there is a massive red galaxy sitting right in the middle of the cluster. 
Then some other process may sometimes bring in other massive red galaxies, but they will 
probably sink into the center over several dynamical times. 

Here, we actually have the ability to see if this scenario is true: we have the 
number of LRGs inside each of these clusters and we know where they are. 
Below we present results on 
the spatial distribution of LRGs in clusters (\S\ref{sec:lrgdist}) and
the halo occupation number (\S\ref{sec:hod}).
and the LRG multiplicity function (\S\ref{sec:mult}),

\subsection{Spatial Distribution of LRGs within Clusters}
\label{sec:lrgdist}

\begin{figure}
\begin{center}
\includegraphics[width=3.0in]{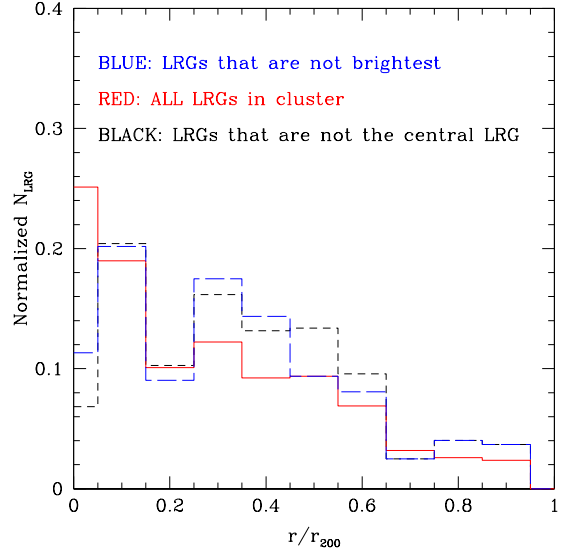}
\end{center}
\caption{The distribution of LRGs in the clusters. The number of LRGs in each bin are normalized 
by dividing the number of LRGs in each bin by the total number of LRGs in all bins.}
\label{fig:dist}
\end{figure}

\begin{figure}
\begin{center}
\includegraphics[width=3.0in]{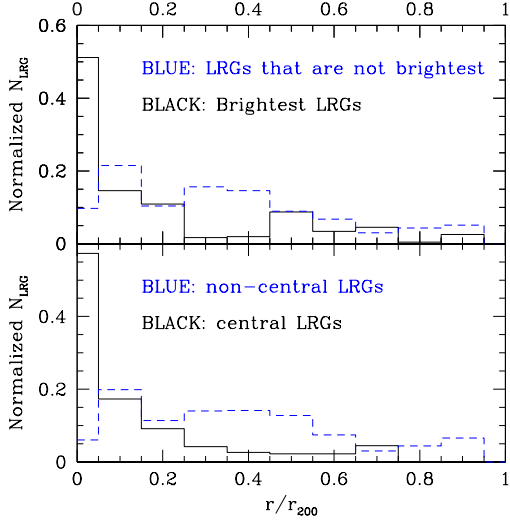}
\end{center}
\caption{(Above) The distribution of brightest LRGs (BLRGs) and the non-brightest LRGs in the clusters.
As shown above, the BLRGs tend to lie at centers of the clusters, while the ones that 
are not BLRGs have a shallower radial distribution.
(Below) The distribution of most central LRGs and the non-central LRGs in the clusters.
The distribution of the central LRGs are very similar to the BLRGs.} 
\label{fig:bdist}
\end{figure}

We show the spatial distribution of LRGs within the clusters in Fig.~\ref{fig:dist}. 
Previous studies \citep[e.g.][]{jones84,lin04} have shown that 
brightest galaxies  
tend to lie at the center of the clusters. 
Here we test if this is true for the LRGs. 
We plot the distribution of brightest LRGs in each of the cluster alongside 
with their companions in each of the cluster (see Fig.~\ref{fig:bdist}).
One realizes that most ($\sim80\%$) of the brightest LRGs resides within the
inner 20\% of the scaled radius of the cluster. 
Therefore, we are consistent with the picture of having brightest galaxies 
lying at the centers of clusters. However, there is a significant fraction of 
clusters that does not follow this rule.

The question of whether the centers of intracluster gas coincide with the central LRGs 
(defined as the LRG closest to the centroid of the X-ray emitting gas)
is also very important to the understanding of the formation of galaxies.
We investigated the distribution of the central LRGs inside the cluster (see Fig.~\ref{fig:bdist}). 
There are $\sim 20\%$ of the ``central'' LRGs which are not central
at all. This may suggest a few scenarios, one being that the cluster is not relaxed 
enough for the central LRG to sit at the center of the gravitational potential (which 
is supposedly traced by the intracluster light).
The centroiding of the clusters in X-ray is called into question, and we will
address this in \S \ref{sec:system}.

\begin{figure}
\begin{center}
\includegraphics[width=3.0in]{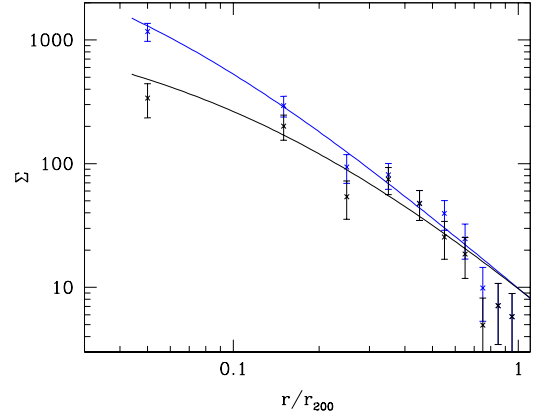}
\end{center}
\caption{The distribution of LRGs in the clusters with the fit to the NFW profile.
Blue line: we fit the surface density of the LRGs (including BLRGs) to a NFW profile and 
get a concentration of $17.5^{+7.1}_{-4.3}$ with
$\chi^{2}=4.29$. Black line: we fit the surface density of the LRGs (excluding BLRGs) to a NFW profile and
get a concentration of $6.0^{+3.2}_{-1.9}$ with $\chi^2=6.6$.}
\label{fig:nfw_p}
\end{figure}

\begin{figure}
\begin{center}
\includegraphics[width=3.0in]{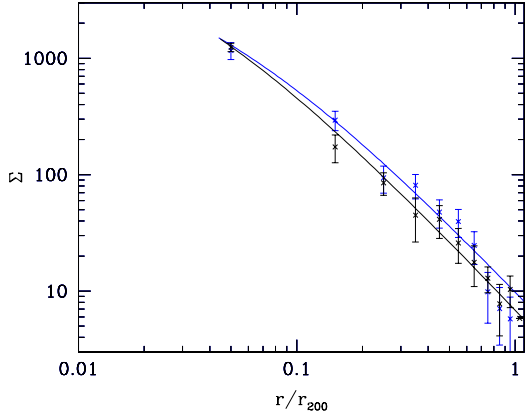}
\end{center}
\caption{The distribution of LRGs in the clusters with the fit to the NFW profile.
Blue line: we fit the surface density of the LRGs (including Brightest LRGs) to a NFW profile and 
get a concentration of $17.5^{+7.1}_{-4.3}$ with
$\chi^{2}=4.29$. Black line shows the fit of the profile of bright galaxies (normalized to this plot) from
\citep{lin06}. }
\label{fig:nfw_r}
\end{figure}

The profile of galaxies in cluster is a key ingredient to the halo model formalism. 
One would like to understand how statistically LRGs populate 
the clusters they are residing. 
We try to fit the NFW profile here to the LRG surface density of stacked clusters in our sample 
and find that the concentration of the surface density to be $17.5^{+7.1}_{-4.3}$ with
$\chi^{2}=4.29$.
We also plot the fitted profile in Fig.~\ref{fig:nfw_p}.
We also fit the NFW profile to LRG surface density of stacked clusters without
the BLRG (Brightest LRG in the cluster), and this gives a concentration of
$6.0^{+3.2}_{-1.9}$ with $\chi^2=6.6$.
Both profiles have very similar concentration as the K-band cluster profile discussed in \citet{lin07} (see
Figure~\ref{fig:nfw_r}).
Errors in $r_{200}$ determination do not affect the fit in any significant fashion as 
demonstrated in \citet{lin07} appendix . 

\subsection{Halo Occupation Number}
\label{sec:hod}

\begin{figure}
\begin{center}
\includegraphics[width=3.0in]{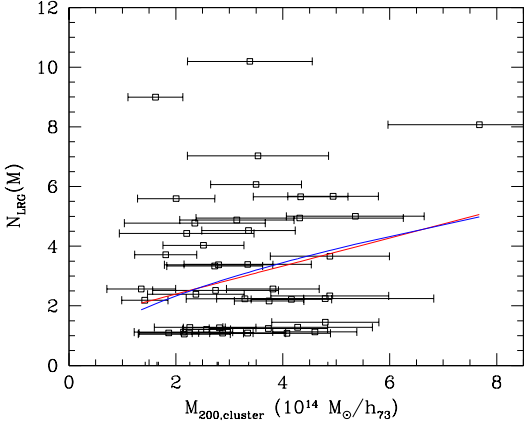}
\end{center}
\caption{The number of LRG per halo $N$ as a function of binned halo mass ($10^{14} h_{73}^{-1} M_{\odot}$),
the fit (red line) is calculated by maximizing the likelihood given a model of $N(M_t) = a*M_t + k$ where $M_t$ is the true measure of $M_{200}$ in $10^{14} h_{73}^{-1} M_{\odot}$, assuming Poisson distribution of $N(M_t)$, where $M_t$ is the true mass of the cluster. We also assume a log-normal distribution for probability of
$P_{ln}(M_t| M_i)$, where $M_i$ is the measured mass. This gives $a=0.470\pm 0.205$ and $k=1.455\pm 0.72$.
The blue line fit is calculated by maximizing the likelihood given a model of $N(M_t) = k*M_t^{a}$ 
and it gives $a=0.560\pm 0.250$ and $k=1.575\pm 0.525$.}
\label{fig:nm}
\end{figure}

As the halo occupation number of the clusters is a key ingredient to the halo model formalism,
we investigate the number of LRGs in these clusters as a function of their masses.
As the size of our sample is not large and the mass estimate o the clusters are accurate to
$30\%-50\%$ only, one will have to be extra cautious in finding a fit for the average number 
of LRGs in the mass range of these clusters. 
We take the following approach, assuming two different models:
\begin{eqnarray}
N(M_t) = a\times M_t + k  \nonumber \\
N(M_t) = k \times M_t^{a}  \nonumber \\
\end{eqnarray}
where $M_t$ is true value of cluster virial mass in 
$10^{14} h_{73}^{-1} M_{\odot}$, a Poisson distribution of $N(M_t)$ and 
two distributions for the probability
finding $M_t$ given $M_i$ where $M_i$ is the measured mass of the i-th cluster (in same units as in $M_t$),
one being Gaussian, the other Log-Normal.
In short, we have the following:
\begin{eqnarray}
L_{tot} = \prod_{i}^{N_c}\int { P(N_i,M_{t,i}|a,k) P (M_{t,i}|M_i) } d M_{t,i} \nonumber \\
\log P(N,M|a,k) = N\times \log (\mu) + const - \mu  \nonumber \\
\log P_g(M_t|M_i) = - (M_t-M_i)^2/(2\sigma_{M}^2) + const  \nonumber \\
\log P_{ln}(M_t|M_i) = -(M_{t,l}-M_{i,l})^2/(2(\sigma_{M_l})^2) + const  \nonumber \\
\label{eq:likelihood}
\end{eqnarray}
where  $\mu = a \times M + k$ or $\mu = k \times M^a$, $M_{t,i}$ stands for 
the $M_t$ for the i-th cluster and $M_{x,l}$ stands for $\log_{10}(M_x)$ 
for the above mentioned form of fit. 
We then maximize the total Likelihood within a grid of resolution $100, 1000, 10000$ 
for both $a$, $k$ and we also vary the size of $d M_{t,i}$ to ensure that 
our results are robust with respect to varying grid size. 
We also try a variety of ranges for both $a$ and $k$ (such as $0<a<20$, $-10<a<10$ and $0<a<3$;
and $0<k<20$, $-20<k<20$ and $0<k<5$), and make sure we get the same answer.
This linear fit with $P_g(M_t|M_i)$ gives $a=0.455\pm 0.215$ and $k=1.605\pm 0.705$ for $68.3\%$ confidence intervals.
This power-law fit with $P_g(M_t|M_i)$ gives $a=0.515\pm 0.245$ and $k=1.725\pm 0.540$ for $68.3\%$ confidence intervals.
See Fig.~\ref{fig:nm} for the data and the fit using $P_{ln}(M_t|M_i)$.

\subsection{$N(M)$ distribituion: Poisson or not?}

Since we assume a Poisson distribution for $N(M_t)$ (hereafter $N$ for simplicity in this section), we test if this is 
a good assumption by looking at $(\langle N^2 \rangle -\langle N \rangle ^2)/ \langle N \rangle $. 
We define $\gamma_{N} = (\langle N^2 \rangle - \langle N \rangle ^2)/ \langle N \rangle $, and since we only have $N(M_i)$, but not $N$ (Number of LRG given the true measure
of cluster mass) 
for each cluster, 
therefore, we have to consider the contribution of scatter from the various systematic effects we mentioned 
in \S\ref{sec:method}: 
\begin{eqnarray}
\langle N_i^2 \rangle_{(M_i)} = \langle N_{int} \rangle^2 + \langle N_{int} \rangle + Y +Z + W \nonumber \\
Y = 2f \langle N_{int}\rangle \langle N\rangle_{(M_i)} \nonumber \\
Z =\int d M_t [P(M_t| M_i) (\langle N \rangle^2_{(M_t)} + V) \nonumber \\
V = \gamma \langle N \rangle_{(M_t)} \nonumber \\
W =  f(1-f)\langle N \rangle_{(M_i)} 
\end{eqnarray}
where $f=f(z_p,z_c,z_s)$ and is defined in \S\ref{sec:method}, $N_{int}$ is the number of interloper as discussed in 
\S\ref{sec:method} and $X_{(M_i)}$ ($X_{(M_t)}$) means the quantity X conditioned on $M_i$ ($M_t$).

Subtracting $\langle N(M_i) \rangle^2_{(M_i)}$ from the equation will reduce to:
\begin{eqnarray}
\langle N_i^2 \rangle _{(M_i)} - \langle N(M_i)\rangle^2_{(M_i)} = PQ+R+S+T \nonumber \\
P = f^2 \gamma + f(1-f) \nonumber \\
Q = \langle N\rangle_{(M_i)} \nonumber \\
R = \langle N_{int}\rangle \nonumber \\
S = f^2 (\int d M_t ([P(M_t| M_i)\langle N \rangle ^2_{(M_t)}) \nonumber \\
T = - f^2 \langle N \rangle ^2_{(M_i)} 
\end{eqnarray}
Note that $\langle N \rangle_{(M_i)}$ = $\int d M_t [P(M_t| M_i) \langle N\rangle_{(M_t)}]$.

\begin{figure}
\begin{center}
\includegraphics[width=3.0in]{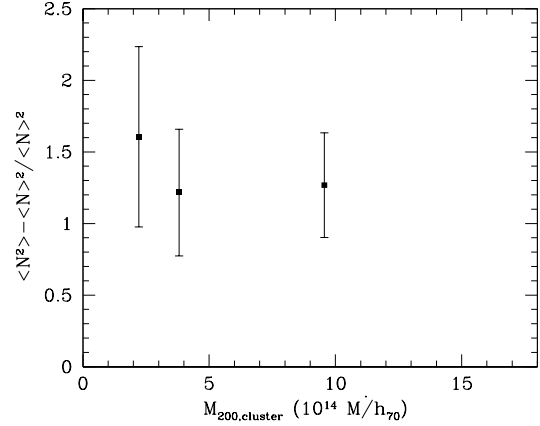}
\end{center}
\caption{To test whether our assumption of a Poisson distribution for $N(M_t)$ is valid, we compute $\langle N(N-1)\rangle/\langle N\rangle^2$ for the combined sample of $400d+Yx$. It does not deviate drastically from being Poisson.}
\label{fig:pois1}
\end{figure}
A Poisson distribution is completely characterized by its first moment, 
$\gamma$ would be 1 if the distribution is completely Poisson. 
We calculate the $\gamma$ from the combined sample of $400d$ and $Yx$ (please refer to \S\ref{sec:system} for a description of $Yx$ sample) sample
and use $\langle N \rangle $ from the fit of $N(M) = k\times (M/10^{14})^{a}$. We bin the cluster such that there are equal number
of clusters in each mass bin (See Figure~\ref{fig:pois1}). 
We find that $\gamma= 1.428\pm 0.351$ and thus
the $N(M_t)$ distribution is consistent with being Poisson.

\begin{figure}
\begin{center}
\includegraphics[width=3.0in]{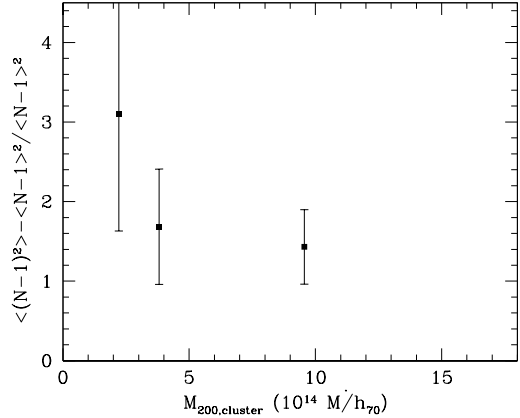}
\end{center}
\caption{To test if the satellite LRG distribution is Poisson, 
we calculate $(\langle(N-1)^2\rangle-\langle(N-1)\rangle^2)/\langle(N-1)\rangle$ for 
the combined sample of $400d+Yx$. It is consistent with being Poisson.}
\label{fig:pois2}
\end{figure}

Furthermore, one important ingredient of halo occupation distribution is the assumption 
of Poisson distribution of the satellite galaxies. We test the assumption here by 
computing  $(\langle (N-1)^2 \rangle - \langle (N-1)\rangle^2)/\langle(N-1)\rangle$ for the $\langle N-1\rangle$ distribution (see Fig. \ref{fig:pois2}) in a similar way as 
we compute $\gamma$ for the $\langle N \rangle$ distribution. 
We find that $\gamma_{N-1} = 1.823 \pm 0.496$ and so
the satellite LRG distribution 
in clusters is also consistent with being Poisson.
However, one should note that it is mathematically impossible for both $N$ and $N-1$ 
to be both exactly Poisson for the same distribution.
The errorbars are calculated via $\sigma(X) = \sqrt((X-\bar{X})^2)/N_c$.

\subsection{LRG Multiplicity Function}
\label{sec:mult}
\begin{figure}
\begin{center}
\includegraphics[width=3in]{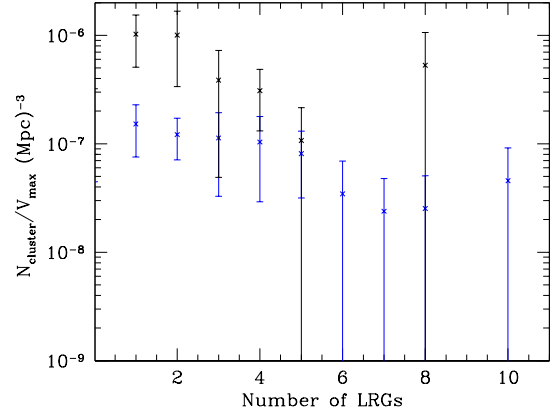}
\end{center}
\caption{The volume weighted multiplicity function of LRGs in these clusters.
Blue (Black) line: the volume weighted multiplicity function for clusters with X-ray 
luminosities $>=$ ($<$) $10^{44} ergs^{-1}$. The variance is calculated by taking the 
sum of $1/(V_{max}^2)$ for each $N_{LRG}$ bin.}
\label{fig:multiplicity}
\end{figure}

Finally we study the multiplicity of LRGs in clusters (Fig.~\ref{fig:multiplicity}). 

We calculate the multiplicity function by  
counting the $1/V_{max}$ weighted number of cluster in each bin.
We compute $V_{max}$ (the comoving search volume of the cluster) by: 

i.We find the flux of the cluster via the following \citet{burenin06}:
\begin{equation}
f = \frac{L}{4 \pi d_L(z)^2} K(z)
\end{equation}
where $L$ is the luminosity of the cluster, $d_L(z)$ is is the cosmological luminosity distance, 
$K(z)$ is the K-correction factor for X-ray clusters (for more details see \citet{burenin06}). 

ii. We find the 
comoving search volume that each cluster with luminosity L can be detected given
by the following:
\begin{equation}
V_{max}(L) = \int_{z=0}^{z=z_c} P_{sel}(f,z) \frac{dV}{dz} dz 
\end{equation}
where $P_{sel}(f,z)$ is the selection efficiency of the 400 square degrees ROSAT PSPC Galaxy Cluster Survey
provided by A. Vikhlinin and R. Burenin (private communication), $dV/dz$ is the cosmological comoving volume per redshift
interval (see \citet{burenin06} for more details).

\section{Systematics}
\label{sec:system}

\subsection{Uncertainties in the Choice of Cluster Radius}



We choose to use $\theta_{200}$ since it is closest to the virial radius of the 
clusters. 
We also look at how the uncertainties of $r_{200}$ will affect our 
results. $r_{200}$ is accurate up to $\sim 10\%$ \citep{reiprich02}. We 
calculate the following to determine the effective number of LRGs 
we are missing due to uncertainties in $r_{200}$:
\begin{equation}
\int_{0}^{r_{200}} \rho(r) dr/ \int_{0}^{1.1r_{200}} \rho(r) dr = 0.95
\end{equation}
and 
\begin{equation}
\int_{0}^{r_{200}} \rho(r) dr/ \int_{0}^{0.9r_{200}} \rho(r) dr =1.06 
\end{equation}

We set the density profile $\rho(r)$ as a NFW profile with concentration of 8 (which is approximately 
what we get when we fit the surface density of the cluster when 
we exclude the BCG). 
This shows that the uncertainties in $\theta_{200}$, thus $r_{200}$ only affect our 
estimation of $N(M)$ at  
the level of $\sim 5\%$.


\subsection{Mass Estimation and Sample Selection}

Cluster mass estimation is crucial in our analysis, as it defines the cluster virial region
to search for member LRGs, and provides a fundamental radius to scale the distance of LRGs
to cluster center. We infer cluster mass through the X-ray luminosity--mass scaling relation
\citep{reiprich02}, which has been shown as a unbiased estimator \citep{reiprich06}.
Compared to other X-ray--based cluster proxies such as temperature and $Y_X$ (the
product of gas mass and temperature, which is proportional to the thermal energy of the
cluster \citealt{kravtsov06}), $L_X$--$M$ correlation shows higher degree of scatter. We therefore seek for
another cluster sample with better measured mass (despite without well-defined selection
criteria).

Recently, \citet{maughan07} have presented a large cluster sample
selected from the {\it Chandra} archive, for which the cluster mass is inferred from $Y_X$,
and the cluster center is inferred from the {\it Chandra} images.
26 of these clusters lie within our SDSS DR5 masks and the redshift range $0.2\le z\le 0.6$.
16 of these 26 clusters do not overlap with our 400d sample 
and we use them to examine the results presented in \S\S\ref{sec:lrgdist} \& \ref{sec:hod}
(hereafter the $Y_X$ sample). 

\begin{figure}
\begin{center}
\includegraphics[width=3.0in]{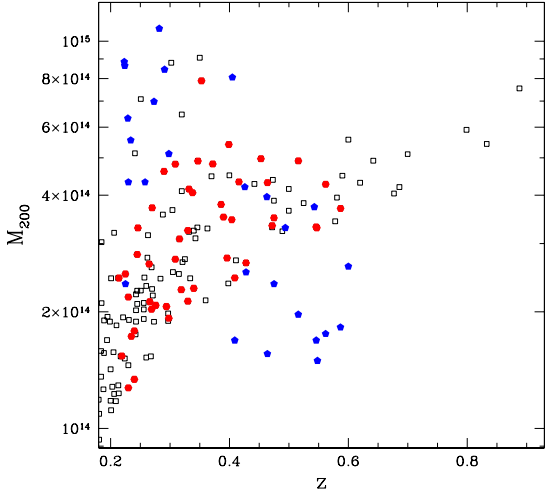}
\end{center}
\caption{Distribution of the clusters on the mass--redshift plane. The squares, red points,
and blue points denote the whole 400d survey sample, the subsample used in this study,
and the $Y_X$ sample, respectively.}
\label{fig:mass_distr}
\end{figure}

\begin{figure}
\begin{center}
\includegraphics[width=3.0in]{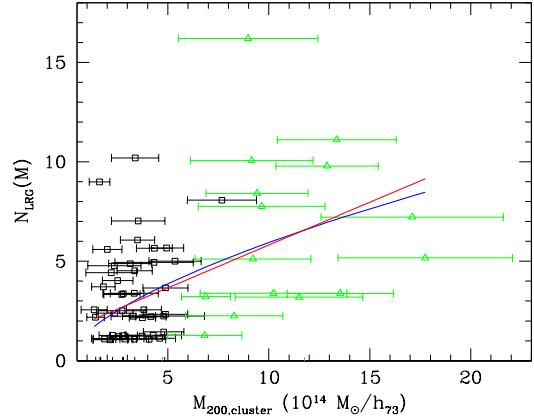}
\end{center}
\caption{Combining the $Y_X$ sample (green triangles), we have a larger mass coverage, thus giving stronger constraints on the slope. We have fairly similar fits between the two different models (linear (red) and power-law (blue)).
The fits are also consistent with the respective fits using only clusters from 400d survey.}
\label{fig:Nmcom}
\end{figure}

Because of the flux-limited nature of the 400d survey, low mass ($\sim 10^{14} M_\odot$) clusters
will be only detected at lower redshifts. In Fig.~\ref{fig:mass_distr} we show the mass distribution
of the whole 400d sample (open squares) and the subsample used in our analysis (red points)
within $0.2\le z\le0.9$. It shows that our sample is a random subsample of the whole 400d sample
Interestingly, at $z \sim 0.3$ the 400d survey clusters
cover a larger range in mass than at other redshifts. It is also clear that the halo occupation numbers
of clusters of lower masses in Fig.~\ref{fig:nm} would be biased to those at $z \le 0.3$.
In Fig.~\ref{fig:mass_distr} we also show the distribution of the $Y_X$ sample. Very curiously,
the distribution of this sample on the mass--redshift space seems to be roughly orthogonal to that
of our 400d sample. Since our results derived from the $Y_X$ sample is consistent with those
based on the 400d sample, we combine the two samples to enhance the mass coverage
(especially for clusters at $z\ge 0.4$) and the statistical signal. We calculated the 
$N(M)$ for the combined sample and assuming power-law model, we have $N(M)= k*(M/10^{14})^{a}$,
where $a = 0.620\pm 0.105$ and $k= 1.425\pm 0.285$, see Fig~\ref{fig:Nmcom}. A more detailed result 
table is shown at Table~2.

\section{Discussion and Summary}
\label{sec:discussion}

\subsection{What is a good mass tracer?}
\label{sec:masstracer}

As clusters are becoming more important
cosmological tools, we need to characterize the masses of clusters more than ever.
``What is a good mass tracer?'' has been a very well-motivated question. 

Here we try to investigate a few options that people have suggested before
as possible solutions:

Fist, as we see earlier in \S\ref{sec:hod}, the mean number of 
LRGs does not trace the masses accurately. 

We quantify this by looking at the scatter 
of the $N_{LRG}-M$ relation by the following quantities in a 3 mass bins:
\begin{eqnarray}
\sigma(ln(N_{LRG})) = \frac{\sqrt(\gamma)}{\sqrt(N_{LRG})} \nonumber \\
\sigma(ln(M)) = \frac{1}{a} \sigma(ln(N_{LRG}))
\end{eqnarray}
where a is as defined in $N(M) = k\times (M/10^{14})^{a}$.
We found that the scatter in $ln(N_{LRG})$ ($ln(M)$) in low, middle and high 
mass bins are $0.332$ ($0.535$) dex (at $M=2.22\times 10^{14} h_{73}^{-1} M_{\odot}$), $0.281$ ($0.452$) dex (at $M=3.81\times 10^{14} h_{73}^{-1} M_{\odot}$) and $0.21$ ($0.340$) dex (at $M=9.56\times 10^{14} h_{73}^{-1} M_{\odot}$) respectively.

\begin{figure}
\begin{center}
\includegraphics[width=3.0in]{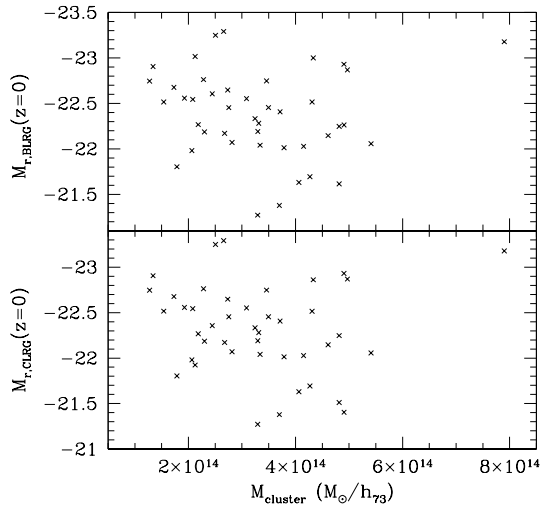}
\end{center}
\caption{(Above) Luminosities of the central LRGs for each cluster.
(Below) Luminosities of the brightest LRGs for each cluster.}
\label{fig:centlumin}
\end{figure}

Second, we look at the luminosities of the central LRG. 
As previous studies suggested in some bands, the brightest cluster galaxies
traces the mass of the cluster \citep{lin04} and that the brightest cluster galaxies tend to be
the central galaxies of the cluster, we look at the relation 
between the luminosities of the brightest LRG in clusters and their X-ray masses.
However, the correlation between the distribution of luminosities of central 
LRG and the masses of the clusters in our sample does not look promising(see Fig.~\ref{fig:centlumin}).

We then look at the correlation between the luminosities of the brightest LRG and their
cluster X-ray masses.
However, it does not seem to be promising either (see Fig.~\ref{fig:centlumin}).
This is also seen in \citet{lin04} when we look at the same mass range and 
when one looks at the correlation between the brightest LRGs and the richness of
the maxBCG catalog \citep{koester07}, there is not a strong correlation for 14,000 clusters 
(R.~Reyes 2007, private communication).
However, there are several caveats that would require further investigations, such 
as the possibility of photo-z failure for the CLRG or BLRG in the clusters and 
possible photometry problem that could destroy the correlation. 
We look into the available spectroscopic data in SDSS and found no 
extra LRGs that are targeted by the SDSS spectroscopy. This rules out the
possible missing LRGs that have $M_r$ of range $\sim -20.8$ (at $z=0.2$) and $\sim -22.5$ (at $z=0.6$). 
Furthermore, as we investigated earlier, only  4 clusters do not have LRGs and 
we find $\sim 70\%$ of BLRGs lie in the central $\sim 20\%$ 
of the virial radius, thus, most clusters do have a LRG at their centers. If we 
are missing Brightest LRGs in centers of clusters, we need to expect the scenario of 
having more than 1 LRG at the central $\sim20\%$ of cluster virial radius to be prevalent. 
This scenario is not supported by the distribution of LRGs as shown in Figure~\ref{fig:bdist}.
Given the caveats and findings here,
we conclude that further work will be needed to make this more quantitative, especially 
to quantify the effect of photometry errors on the correlation.



\subsection{Evolution of Massive Galaxies}
\label{sec:evolution}

Insights into the evolution of massive galaxies in clusters may be
gained by comparing
some of the results presented in \S\ref{sec:Analysis} with the
properties of cluster galaxies
in the local Universe. For consistency with our LRG selection, we select
nearby cluster galaxies by
the requirement that they are one magnitude more luminous than the
characteristic magnitude
($M_*-1$).

We first examine the spatial distribution of massive galaxies within
clusters.
\citet{lin04}, with a large sample of clusters at $z<0.2$, find
that luminous cluster
galaxies ($M_K\le -25$) follow an NFW profile with concentration of
$18.2$ ($5.8$), when the brightest
cluster galaxy is included (excluded). This result is in very good
agreement with our finding
in \S\ref{sec:lrgdist}.

The second comparison is made with the halo occupation number. We
construct the occupation
number for $M_K\le -25$ with the $z<0.1$ cluster sample presented in \citet{lin06}. No
color selection analogous to that presented in \S\ref{sec:LRGdata} is used.
However, for such a luminous
magnitude range, the contamination of blue galaxies should be
minimal. Nevertheless,
we should regard the mean occupation number $\bar{N}$ thus obtained as a
upper limit.
We find that the nearby $\bar{N}$--$M$ relation is similar to that shown in \S\ref{sec:hod}.

Taken at face value, these comparisons seem to suggest that there is not
much evolution in
the massive galaxy populations between $z\sim 0.5$ and $z\approx 0$. The
occupation number
comparison basically suggests that the shape of the luminosity function
is similar in clusters
at these two epochs, after the passive evolution has been taken into
account.
This is consistent with several previous studies, both for cluster
galaxies \citep{lin06,andreon06,depropris07,muzzin07}
and the field population e.g.~\citep{wake06,brown07}.

However, evidence for mergers that produce massive galaxies has been
found (\citealt{vandokkum05,bell06}; some other references). In the $\Lambda$CDM model,
formation of massive
objects through mergers of less massive ones is a generic feature. To
reconcile the apparent no-evolution of the aforementioned bulk properties
with this picture, we suggest two considerations.
(1) Irrespective of the role of mergers in the
formation and evolution of the LRGs, their spatial distribution seems to
be similar out to $z\sim 0.5$.
This is similar to the ``attractor'' hypothesis of \citet{gao04}.
(2) The degree of evolution, be it an increase in the number of LRGs due
to mergers of the host halo
with less massive halos, or a decrease due to dynamical processes
(e.g.~tidal disruption, mergers),
would be seen more clearly  through (Monte Carlo) simulations where the
merger history of the
halos is fully followed. In a companion paper such an approach is
adopted to infer the merger rate of LRGs \citep{conroy07}.

\subsection{Summary}

We investigate statistical properties of LRGs in a sample of X-ray
selected galaxy clusters at intermediate redshift ($0.2\le z\le0.6$).
The LRGs are selected based on carefully designed color criteria, and
the cluster membership is assessed via photometric redshift.
We put constraints 
on spatial distributions of LRG within clusters, namely the radial 
distribution. 
We find that the distribution of brightest LRGs in cluster 
to be concentrated as discussed in previous studies \citep{jones84,lin04}.
We also find that the radial distribution can be fit by a NFW profile
with a concentration of 
$17.5^{+7.1}_{-4.3}$ with
$\chi^{2}=4.29$ when we include the brightest LRG.
When we do not include the brightest LRG, we find concentration of 
$6.0^{+3.2}_{-1.9}$ with $\chi^2=6.6$.
Considering the sample size and mass errors on our sample, we 
use the maximum likelihood method to find the best fit 
parameters for halo occupation distribution ($N(M)$). 
The result depends on what kind of models we adopt, but are 
fairly insensitive to what model we use, results are shown in Table~2. 

Uncertainties in photometric redshifts are taken into account
by including different possible effects such as interlopers and
missing LRGs due to errors in photometric redshifts (see \S\ref{sec:Analysis}).
We estimate that the errors in cluster radius can only
contribute to our uncertatinty in $N(M)$ at the level of $\sim 5\%$.
Errors in mass estimation are fully taken into account throughout the 
analysis. We also employ an independent sample of better measured  
masses ($Y_X$ sample) to test the mass estimation of our sample.
However, we do implicitly assume that the scatter of $M-L_X$ relation
does not correlate with $N(M)$ during the analysis.
The result we derive from a combined analysis of both sample 
on $N(M)$ is consistent with using our sample alone (see Table~2).
We also find that there are no obvious good mass tracer as 
we look at different correlations between various quantities of clusters and 
their galaxies. 
Last, we discuss the evolution of massive galaxies from different perspectives.
We conclude that it would be important to study low-z LRG population to better constrain the 
evolution of the population \citep{ho07}.

\noindent
\clearpage
\begin{table}
{\footnotesize
\begin{center}
\label{tab:t1}
{\footnotesize
\begin{tabular}{llllllll}
\hline
Name & RA   &  DEC    &  redshift   &  $M_{200}$  & $\theta_{200}$ & LRG  & LRG count \\
 &     (deg)   & (deg)   &  &  ($10^{14} M_{\odot}h_{73}^{-1}$)  & (arcmin) & count & corrected \\
\hline 
20  &  29.8258  & 0.5025 & 0.386 & 3.815 &  4.0608 & 2 & 2.564 \\
36  &  46.7695  & -6.4808 & 0.347 & 4.879 &  4.8054 & 2 & 2.336  \\
80  &  122.4208  & 28.1994 & 0.399 & 5.357 &  4.428 & 4 & 5.003  \\
86  &  132.2975  & 37.5230 & 0.240 & 1.419 &  4.326 & 2 & 2.191  \\
88  &  133.3058  & 57.9955 & 0.475 & 3.536 &  3.3564 & 5 & 7.027  \\
99  &  149.0116  & 41.1188 & 0.587 & 3.731 &  2.895 & 1 & 1.243\\
100  &  149.5541  & 55.2683 & 0.214 & 2.516 &  5.7744 & 3 & 4.031  \\
101  &  149.5804  & 47.0380 & 0.390 & 3.552 &  3.933 & 0 & -0.0116  \\
103  &  150.7687  & 32.8933 & 0.416 & 4.334 &  3.9906 & 4 & 5.659  \\
107  &  152.8558  & 54.8350 & 0.294 & 2.142 &  4.185 & 1 & 1.112  \\
108  &  153.3658  & -1.6116 & 0.276 & 2.157&  4.4214 & 1 & 1.052 \\
110  &  154.5037  & 21.9097 & 0.240 & 1.867 &  4.74 & 1 & 1.093  \\
111  &  156.7945  & 39.1350 & 0.338 & 4.079 &  4.6248 & 1 & 1.083  \\
121  &  169.3754  & 17.7458 & 0.547 & 3.343 &  2.949 & 3 & 3.389  \\
123  &  170.2429  & 23.4427 & 0.562 & 4.277 &  3.1344 & 1 & 1.281 \\
124  &  170.7941  & 14.1611 & 0.340 & 2.375 &  3.843 & 2 & 2.388  \\
134  &  178.1487  & 37.5461 & 0.230 & 2.260 &  5.238 & 1 & 1.279 \\
136  &  180.0320  & 68.1519 & 0.265 & 2.726 &  4.9458 & 3 & 3.334  \\
137  &  180.2062  & -3.4583 & 0.396 & 2.818 &  3.5964 & 1 & 1.265  \\
142  &  183.0800  & 27.5538 & 0.353 & 7.677 &  5.5116 & 7 & 8.071  \\
144  &  183.3933  & 2.8991 & 0.409 & 2.518&  3.3756 & 0 & -0.008  \\
145  &  184.0825  & 26.5558 & 0.428 & 2.742 &  3.3492 & 2 & 2.517\\
146  &  184.4320  & 47.4872 & 0.270 & 3.743 &  5.412 & 2 & 2.164  \\
150  &  185.5079  & 27.1552 & 0.472 & 3.384 &  3.324 & 7 & 10.194  \\
163  &  193.2695  & 62.8027 & 0.235 & 1.810 &  4.7766 & 3 & 3.714  \\
166  &  197.1370  & 53.7041 & 0.330 & 2.208 &  3.8436 & 0 & -0.007 \\
167  &  197.8029  & 32.4827 & 0.245 & 2.876 &  5.379 & 1 & 1.089 \\
168  &  198.0808  & 39.0161 & 0.404 & 3.500 &  3.8046 & 5 & 6.069 \\
172  &  202.8791  & 62.6400 & 0.219 & 1.619 &  4.8876 & 7 & 8.992  \\
175  &  204.7091  & 38.8550 & 0.246 & 3.342 &  5.6358 & 1 & 1.087 \\
181  &  208.5695  & -2.3627 & 0.546 & 3.360 &  2.958 & 4 & 4.525  \\
184  &  212.5558  & 59.7105 & 0.316 & 3.140 &  4.479 & 4 & 4.879 \\
185  &  212.5662  & 59.6408 & 0.319 & 2.356 &  4.0386 & 4 & 4.775  \\
188  &  214.6300  & 25.1797 & 0.290 & 4.599 &  5.4612 & 1 & 1.127  \\
198  &  231.1679  & 9.9597 & 0.516 & 4.881 &  3.5016 & 3 & 3.662  \\
202  &  243.5479  & 34.4236 & 0.269 & 2.110 &  4.4844 & 0 & -0.006 \\
208  &  250.4679  & 40.0247 & 0.464 & 4.316 &  3.654 & 3 & 4.946  \\
209  &  254.6412  & 34.5022 & 0.330 & 3.293 &  4.3914 & 2 & 2.237  \\
210  &  255.1779  & 64.2161 & 0.225 & 2.576 &  5.5746 & 1 & 1.209  \\
211  &  255.3441  & 64.2358 & 0.453 & 4.940 &  3.8952 & 4 & 5.670  \\
212  &  260.7245  & 41.0916 & 0.309 & 2.799 &  4.3908 & 3 & 3.375  \\
s9  &  209.89166 & 62.3169 & 0.332 & 4.162 &  4.725 & 2 & 2.217 \\
s11  &  221.0266  & 63.7483 & 0.298 & 2.006 &  4.0488 & 5 & 5.595  \\
s12  &  225.0108  & 22.5680 & 0.230 & 1.353 &  4.4142 & 2 & 2.564  \\
s13  &  228.5916  & 36.6061 & 0.372 & 4.795 &  4.5156 & 1 & 1.448  \\
s14  &  234.1470  & 1.5556 & 0.309 & 4.793 &  5.253 & 2 & 2.244  \\
s17  &  236.8350 & 20.9502 & 0.266 & 2.201 &  4.5912 & 4 & 4.426  \\
\hline	 	
\end{tabular}
}
\end{center}
\caption{Basic parameters of our cluster sample. Naming scheme follows the cluster number as given in table 4 of  \citet{burenin06} those start with s are from table 5 of \citet{burenin06}, and are not part of the main sample of serendipitous 400d survey.}
}
\end{table}
\clearpage

\begin{table}
\label{tab:nm_com}
{
\begin{center}
{
\begin{tabular}{llll}
\hline
$N(M)$  &  Data       &  Poisson+Gaussian     &  Poisson+log-Normal  \\
\hline
$a*M+k$ &  400d       &  $a=0.455\pm 0.215$     & $a=0.470\pm 0.205$ \\
        &             &  $k=1.605\pm 0.705$     & $k=1.455\pm 0.720$ \\
$a*M+k$ &  400d+$Y_X$ &  $a=0.460\pm 0.090$     & $a=0.430\pm 0.085$ \\
        &             &  $k=1.500\pm 0.435$     & $k=1.515\pm 0.435$ \\
$k*M^{a}$ &400d       & $a=0.515\pm 0.245$      & $a=0.560\pm0.250 $ \\
        &             &  $k=1.725\pm 0.540$     & $k=1.575\pm 0.525$ \\
$k*M^{a}$&400d+$Y_X$  & $a=0.625\pm 0.110$       & $a=0.620\pm 0.105 $ \\
     &                &  $k=1.470\pm 0.285$      & $k=1.425\pm0.285 $ \\
\hline
\end{tabular}
}
\end{center}
\caption{Results of maximizing likelihood assuming different parameters. This table describes the model of the $N(M)$ in the first column, dataset we use in the "Data" column, results of maximizing likelihood by
assuming Poisson distribution for $N(M)$ (both column 3 and 4), Gaussian and Log-Normal distribution for the cluster mass
distribution for column 3 and 4 respectively.}
}
\end{table}

{Acknowledgments:} 
We thank Nikhil Padmanabhan, Charlie Conroy, Jim Gunn, Zheng Zheng, 
and Thomas Reiprich for helpful discussions, and R.A. Burenin and A. Vikhlinin for help with the selection function of the
400d survey.
YTL acknowledges support from the Princeton-Catolica Fellowship, NSF PIRE 
grant OISE-0530095, and FONDAP-Andes. 
Funding for the SDSS and SDSS-II has been provided by the Alfred
P. Sloan Foundation, the Participating Institutions, the National
Science Foundation, the U.S. Department of Energy, the National
Aeronautics and Space Administration, the Japanese Monbukagakusho, the
Max Planck Society, and the Higher Education Funding Council for
England. The SDSS Web Site is http://www.sdss.org/.

The SDSS is managed by the Astrophysical Research Consortium for the
Participating Institutions. The Participating Institutions are the
American Museum of Natural History, Astrophysical Institute Potsdam,
University of Basel, University of Cambridge, Case Western Reserve
University, University of Chicago, Drexel University, Fermilab, the
Institute for Advanced Study, the Japan Participation Group, Johns
Hopkins University, the Joint Institute for Nuclear Astrophysics, the
Kavli Institute for Particle Astrophysics and Cosmology, the Korean
Scientist Group, the Chinese Academy of Sciences (LAMOST), Los Alamos
National Laboratory, the Max-Planck-Institute for Astronomy (MPIA),
the Max-Planck-Institute for Astrophysics (MPA), New Mexico State
University, Ohio State University, University of Pittsburgh,
University of Portsmouth, Princeton University, the United States
Naval Observatory, and the University of Washington.

\bibliography{lrg_mn_2}

\end{document}